\documentclass[Crown,sageh,times]{sagej}

\usepackage{moreverb,url}
\usepackage{float}
\usepackage[colorlinks,bookmarksopen,bookmarksnumbered,citecolor=red,urlcolor=red]{hyperref}

\newcommand\BibTeX{{\rmfamily B\kern-.05em \textsc{i\kern-.025em b}\kern-.08em
T\kern-.1667em\lower.7ex\hbox{E}\kern-.125emX}}

\begin{document}

\runninghead{Liang, Hutson \& Keyes}

\title{Surveillance, Stigma \& Sociotechnical Design for HIV}

\author{Calvin Liang\affilnum{1}\affilnum{*}, Jevan Hutson\affilnum{2}\affilnum{*}, and Os Keyes\affilnum{1}}

\affiliation{\affilnum{*} Liang \& Hutson should be treated and understood as \textit{joint first authors} of this publication\\
\affilnum{1}Department of Human Centered Design \& Engineering, University of Washington\\
\affilnum{2}School of Law, University of Washington}

\corrauth{Calvin Liang, Department of Human Centered Design \& Engineering, University of Washington, 428 Sieg Hall, Box 352315, Seattle, WA, 98195-2315, United States}

\email{cliang02@uw.edu}

\begin{abstract}
Online dating and hookup platforms have fundamentally changed people’s day-to-day practices of sex and love---but exist in tension with older social and medicolegal norms. This is particularly the case for people with HIV, who are frequently stigmatized, surveilled, ostracized and incarcerated because of their status. Efforts to make intimate platforms “work” for HIV frequently focus on user-to-user interactions and disclosure of one’s HIV status but elide both the structural forces at work in regulating sex and the involvement of the state in queer lives. In an effort to foreground these forces and this involvement, we analyze the approaches that intimate platforms have taken in designing for HIV disclosure through a content analysis of 49 current platforms. We argue that the implicit reinforcement of stereotypes about who HIV is or is not a concern for, along with the failure to consider state practices when designing for data disclosure, opens up serious risks for HIV-positive and otherwise marginalized people. While we have no panacea for the tension between disclosure and risk, we point to bottom-up, communal, and queer approaches to design as a way of potentially making that tension easier to safely navigate. 
\end{abstract}

\keywords{HIV, privacy, online dating, sexuality, stigma, surveillance}

\maketitle

\section{Introduction}
\begin{quotation}
\textit{``AIDS is essentially a crisis of governance, of what governments do and do not do, to and for their people---we have the drugs to treat HIV infection, we have the tools to confront the risks that drive HIV transmission and prevent infection itself---what we don’t have is national political will necessary to scale-up our response. We have demanded too little from our leaders, excused far too much.''}\\
\\Gregg Gonsalves, speech at the 2006 Toronto AIDS Conference
\end{quotation}

\begin{quotation}
\textit{``Design is inherently about change---not just in the creation of new material artifacts, but in the ways that new technological objects afford new practices, social habits, and ways of living and interacting.''}\\
\\ ``Social Justice-Oriented Interaction Design: Outlining Key Design Strategies and Commitments''
\end{quotation}

Living and loving with HIV is a complicated task. HIV status and the stigma attached to it exists within a complex interplay of social norms and medicolegal infrastructures. The medicolegal history of HIV begins the moment that HIV and AIDS emerged, constituting a mix of medically-justified legal norms and legally-enforced medical requirements. The criminal justice and public health systems of modern states demarcated people living with HIV as a uniquely dangerous population, ``one that needed to be sought out, tracked down, tested, reported, listed, tagged, monitored, regulated, and, increasingly, criminalized''\citep[p.349]{halperin2017war}.

The immediate policy response in the United States imposed significant criminal and civil liability upon people living with HIV~\citep{hoppe2017punishing,harsono2017criminalization,gostin1999law,sykes2016cruel,Thrasher2015,galletly2014criminal,gagnon2012toward,pollard2006sex}. Between 1986-2019, HIV-specific criminal laws and sentence enhancements applicable to people living with HIV have been enacted in 34 states and two U.S. territories~\citep{chiv2019,lehman2014prevalence}. Since 1986, these laws have criminalized nondisclosure of HIV and engagement in ``risky'' behaviors such as sexual activity, exposure to bodily fluids, needle sharing, sex work, blood/organ/semen donation, and, in a variety of instances, behaviors posing little, if any, risk of HIV transmission~\citep{cdc2019a,chiv2019}.

Despite claiming medical legitimacy for this punitive approach, researchers have long understood that the criminalization of HIV transmission was instead fueled by the associations between HIV and the gay community and communities of color~\citep{hoppe2017punishing,gallo2006reflection,johnson1992silence,banks1989women} at a time when consensual sex between same-sex partners was a criminal offense in twenty-two states and over 61 percent of American evangelicals and 50 percent of non-evangelicals agreed with the statement ``I sometimes think AIDS is a punishment for the decline in moral standards''~\citep{Gallup1987}.

A significant body of empirical social science work documents the harmful effects HIV laws have had on the lives of people living with HIV~\citep{barre2018expert,harsono2017criminalization,sweeney2017association,adam2014impacts}. HIV criminalization both reinforces and magnifies HIV-related stigma and discrimination, reduces the willingness of persons at risk for HIV to get tested or seek care, and imperils demographic health collection of information~\citep{harsono2017criminalization,burris2008case,galletly2006conflicting,Elliot2002}. A survey of over 2,000 people living with HIV in the U.S. revealed that at least 25 percent of respondents knew one or more individuals who were afraid to get tested for fear of facing criminalization~\citep{Sero2012}. HIV criminalization also ignores the reality that successful antiretroviral therapy can render the level of the virus to undetectable, which, according to the National Institute of Health, means that HIV is then untransmittable~\citep{eisinger2019hiv}.

While HIV transmission was criminalized, other tools of control---in the form of surveillance---arose and were enforced. Early policy responses to HIV centered on overt surveillance and ostracism of those infected and perceived to be at risk~\citep{fortin1995aids}. This surveillance generally consists of disease reporting, sexual contact tracing, and data collection of people who have been diagnosed with HIV~\citep{fan2011sex,fan2012decentralizing,ward2005partner,ward2014partner}. The Center for Disease Control, for example, collects HIV data based on confidential name-based reporting laws implemented in all 50 states as of April 2008~\citep{cdc2019b}.

HIV surveillance (and Sexually Transmitted Infection surveillance more broadly) centralizes information and power in the state~\cite{fairchild2007searching,fan2012decentralizing}; because HIV intervention and surveillance is generally concentrated in lower income communities and health settings~\citep{mccree2010contribution}, the most socially and economically marginalized communities bear the heaviest burden of HIV surveillance and its downstream consequences~\citep{miller2004prevalence,banks1989women}. There is a long racialized history of HIV, one that, in combination with the background racism of the United States, has led to the systemic undertreatment and under-consideration of communities of color~\citep{ford2007black,Anonymous2000,johnson1992silence}.

This infrastructure of surveillance in turn reinforces the stigma of HIV, which has dramatic consequences for the likelihood of unwanted disclosure, access to care, psychiatric wellbeing, housing and employment discrimination, and, consequently, quality (or probability) of life~\citep{lazarus2016beyond,mahajan2008stigma}. Coupled with the overarching stigma of HIV and its criminalization in various contexts, HIV surveillance offers a tool through which the state can identify citizens to be punished.

In the era of ``big data'' and ubiquitous surveillance capitalism~\citep{zuboff2019age}---the private monetization of information about reality---HIV surveillance is not just in the hands of the state, but also in the hands of private organizations and individuals. In the context of widespread state surveillance and control and ongoing stigmatization of HIV, this opens yet more possibilities for harm through enabling the selling and redistribution of HIV status information, without the user’s meaningful consent, to parties who may themselves engage in discrimination or direct violence. 

Many online platforms---including, as we trace out below, dating platforms---constitute not just spaces for the purposes outlined in their marketing materials but also tools for the police in tracing HIV status and criminalized behavior. In recent years, police have used technology to conduct Internet-based investigations for a similar purpose (``Man with HIV Arrested for Seeking Sex on Social Media'', ~\citep{Poz2015}). Police now go undercover on websites and dating apps by creating fake identities online~\citep{semitsu2011facebook}, and local law enforcement agencies and federal agencies increasingly employ these tactics in online investigations~\citep{lichtblau2014more}.

Legal and public health scholars and advocates continue to call for a paradigm shift in managing HIV that leaves behind historical responses like surveillance, ostracism, and incarceration and accounts for the rise of the Internet and mobile technology and their impact on sexual attitudes and behaviors~\citep{lehman2014prevalence,mccallum2014criminalizing,fan2011sex}. Since the criminalization of HIV, intimate platforms have become vital structures through which millions of people access the opportunity to engage in reciprocal romantic and sexual relationships~\citep{hutson2018debiasing,taylor2017social,rosenfeld2012searching}. By designing infrastructures for intimate affiliation, intimate platforms wield unmatched structural power to shape who meets whom and how within dating and sexual platforms~\citep{hutson2018debiasing,levy2017designing,emens2008intimate,robinson2007structural}. These platforms frame the circumstances within which users understand each other as prospective romantic or sexual partners and shape social norms, sexual scripts, and relative advantages among users~\citep{hardy2017constructing,kannabiran2012designing}.

The design of intimate platforms provides opportunities to explore new ways of managing HIV that reduce the concentration of power and information in the state \citep{fan2012decentralizing}. Through the role that platform design plays in shaping cultural norms, which has been identified as a more effective way of achieving HIV transmission prevention than flexing the punitive and surveillant arms of the state~\citep{sunstein1996social}, intimate platform design provides opportunities to explore new ways of managing HIV \citep{fan2012decentralizing}. Indeed, a meta-analysis of HIV prevention efforts found that strategies that intervene in social meaning by shaping social norms, cultural practices, and individual attitudes were more effective in empowering behavioral change than appeals to fear~\citep{albarracin2005test}.

However, designing intimate platforms to account for HIV also presents serious challenges for social computing researchers and human-computer interaction (HCI) designers. As Handel and Shklovski pointed out: ``The minutiae of design decisions around profile options deserves particular attention because even the smallest changes can result in substantial differences for user interactions''~\citep{handel2012disclosure}. In addition to concerns around how to best design for HIV, platforms, Grindr in particular, have already come under fire for sharing user HIV information with third parties~\citep{Singer2018}. Moreover, designing intimate platforms to unburden the risks of extant criminal and civil sexual regulations runs the serious risk of re-entrenching the status quo and its incumbent inequalities and power relations~\citep{bardzell2010feminist}. While designing for HIV presents opportunities to redress stigma and harm, researchers in HCI must understand that ``[i]t is not enough to have good intentions...[we] must ground [our] efforts in clear political commitments and rigorous evaluations of the likely consequences''~\citep[p.46]{green2018data}.

From this comes the recognition that social computing designers and researchers seeking to design for disclosure cannot afford to ignore the ways that the lived experiences of people living with HIV are shaped by structural forces and, particularly, the reality of HIV criminalization and the State’s role in conducting STD surveillance. Platforms, after all, do not exist in a separate sphere from material reality: a redesign that eases HIV disclosure from user-to-user might also involve the storing of disclosure data by the platform---data that can then be accessed, requisitioned, and co-opted by arms of the state. In line with Jackson et al.’s call for the social computing community to address the structural and lived consequences of law and policy that ``establish the very terrain on which design and practice can be conceived, articulated, and imagined---and upon which battles of accountability are inevitably waged''~\citep[p.596]{jackson2014policy}, we wish to undertake a critical investigation of HIV disclosure in dating and hookup platforms. This involves not just investigating the implications of disclosure in a person-to-person sense, but also how platform design is shaped by legal and administrative regulation and how the risks of disclosure might open users up to systems of surveillance, stigma, and criminalization. We do so by using a range of platforms in an effort to gain a wide view, and to practice prefigurative politics---minimizing our assumptions about the ``type'' of people at risk of HIV infection and/or surveillance.

To do this, we analyze platforms’ consequences for HIV through the lens of user-to-user interactions, exploring the ways that design renders users visible and vulnerable to wider carceral and surveillance infrastructures, and the way that design shapes (and is shaped) by HIV’s legal status. We ground our discussion in a content analysis of 49 popular, mobile dating and hookup platforms, coding for design and policy choices related to HIV disclosure, prevention, destigmatization, surveillance, privacy, and criminalization. Through this, we reveal that many platforms fail to account for HIV, and of those that do, many neglect to attend to the downstream consequences of HIV disclosure and the data produced by it, while exacerbating the social, racial, and class stereotypes associated with the condition.

As scholars and designers consider how platform design might aid HIV prevention and destigmatization~\citep{hutson2018debiasing,wohlfeiler2013can,rosser2011future}, we aim to grapple with the structural and ethical implications of designing for HIV, particularly how intimate platform design might aid and abet the decriminalization and surveillance of HIV~\citep{sykes2016cruel,kazatchkine2015ending,perone2013punitive,gagnon2012toward,jurgens2009ten}. Drawing on principles from social justice-oriented design to investigate controversies and design possibilities in intimate platforms, we attempt to articulate an approach to intimate platform design that not only works to reduce the stigma of user disclosure, but also works to contest historic and present power imbalances and injustices between users, platforms, and the state.

\section{Methodology}

Using a directed content analysis~\citep{hsieh2005three}), we reviewed 49 existing mobile dating and hookup platforms. Content analyses have proven effective in understanding platform design and governance and the ways design practices mediate user-to-user bias and discrimination~\citep{levy2017designing,hutson2018debiasing}. We set out to capture a landscape of popular platforms and selected the first 49 dating and hook up platforms in the top 200 grossing social networking applications in the United States on the iOS App Store in March of 2019. \hyperref[T1]{Table 1} lists the platforms selected in alphabetical order.

\begin{table}[h]
\small\sf\centering
\caption{Platforms that formed part of our analysis.\label{T1}}
\begin{tabular}{ccccc}
\toprule
Adam4Adam&Cupid&Hud&MR X&Surge\\
Ashley Madison&DOWN&Jack'd&OKCupid&Tagged\\
Badoo&eHarmony&Jaumo&OurTime&Taimi\\
BlackPeopleMeet&Grindr&JDate&PlanetRomeo&Tinder\\
Blendr&GROWLr&LOVOO&Plenty of Fish&Uniform Dating\\
Bumble&Happn&Mamba&Qeep&Waplog\\
Chappy&HER&Match&SayHi&Wild\\
Clover&Hinge&MeetMe&SCRUFF&Xtremboy\\
Coffee Meets Bagel&Hornet&Mingle2&SinglesAroundMe&Zoosk\\
Compatible Partners&Hot or Not&MocoSpace&Skout\\
\bottomrule
\end{tabular}
\end{table}

Utilizing the walkthrough method~\citep{light2018walkthrough}, we explored each platform’s HIV-related user experience. We examined design features on each of these platforms, systematically documenting design choices, policies, and informational interventions that mediate HIV. Building upon previous work around intimate platforms and HIV, we coded each of the 49 intimate platforms based on the following dimensions:

\begin{itemize}
    \item \textbf{\textit{Prevention}}
        \begin{enumerate}
            \item Whether the app allows same-sex connections
            \item Whether a user can disclose HIV/Sexually Transmitted Infection (STI) status~\citep{warner2018privacy}
            \item If they can disclose, what are the options?~\citep{warner2018privacy}
            \item Whether a user can search for or filter out users with HIV / STIs~\citep{hutson2018debiasing}
            \item Whether the platforms provide informational interventions with respect to HIV/STI prevention~\citep{wang2019using}
        \end{enumerate}
    \item \textbf{\textit{Stigma Reduction}}
    \begin{enumerate}
        \item Whether a user can identify as having HIV/STI (e.g., ``Poz'' etc.)
        \item Whether a user can indicate interest in or acceptance of people living with HIV/STIs (e.g. outward presentation, separate from filtering, not simply via profile text)~\citep{hutson2018debiasing}
    \end{enumerate}
    \item \textbf{\textit{Policies}}
    \begin{enumerate}
        \item Whether the platform engages HIV/STIs in their policies (terms of service, privacy, and community policies, etc.)~\citep{jackson2014policy}
    \end{enumerate}
    \end{itemize}

For ethical reasons, we did not interact with other users, only observed features, and deleted our accounts once data were collected when possible (not all platforms allowed for account deletion). The design and policy choices described and discussed below are not intended as an endorsement of any particular design intervention for managing HIV. Rather, we aim to capture the various ways intimate platforms currently manage and mediate HIV among users and how those choices map onto extant legal and surveillant infrastructures. Additionally, we highlight two limitations in how we chose which platforms to analyze. First, it is possible for a hook-up platform to not have an accompanying mobile app, meaning our selection of platforms from the iOS app store will have invariably missed website-based platforms. Second, we may have overlooked platforms that are more niche or community-specific, yet not as popular in the broader platform marketplace (i.e. not within the top grossing platforms).

\section{Findings}
\begin{figure}
\centering
\includegraphics[scale=0.25]{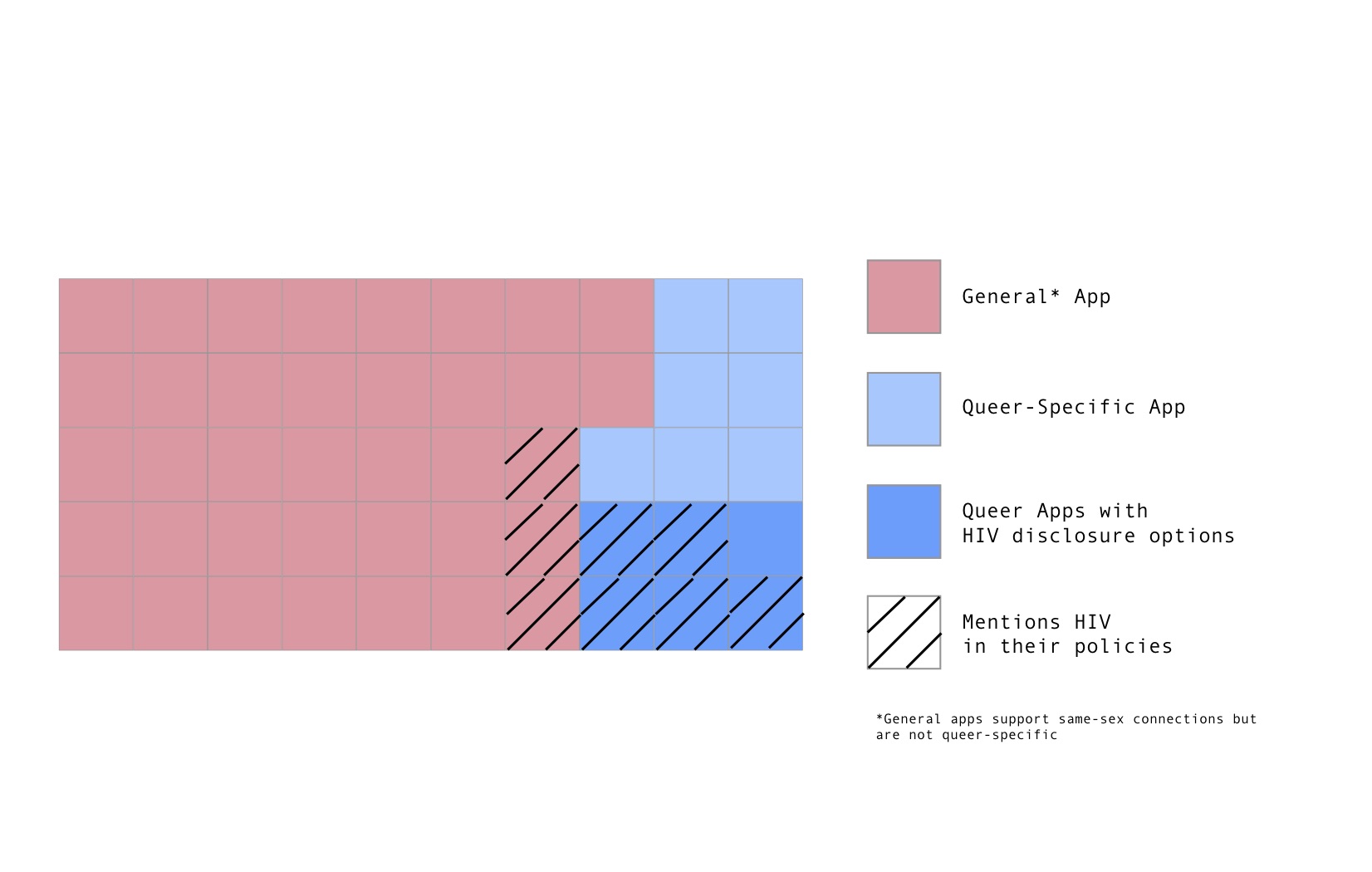}
\caption{A visual representation of our content analysis\label{F1}}
\end{figure}

\subsection{Design Features}

\begin{table}[h]
\small\sf\centering
\caption{Disclosure options found in each platform's design.\label{T2}}
\begin{tabular}{ll}
\toprule
Platform&Disclosure options\\
\midrule
Grindr&Do Not Show, Negative, Negative on PrEP, Positive, Positive Undetectable\\
Hornet&Do Not Show, Not Sure, Negative, Positive, Neg on PrEP, Positive Undetectable\\
GROWLr&HIV-, HIV+, Undetectable, On PrEP, Use Condoms, Drug Free\\
Adam4Adam&Negative, Positive, Positive Undetectable, I Don't Know\\
Mr. X&No Answer, Negative, Positive, Don't Know, Undetectable\\
Scruff&Poz, Treatment as Prevention\\
\bottomrule
\end{tabular}
\end{table}

\begin{figure}
\centering
\includegraphics[scale=0.5]{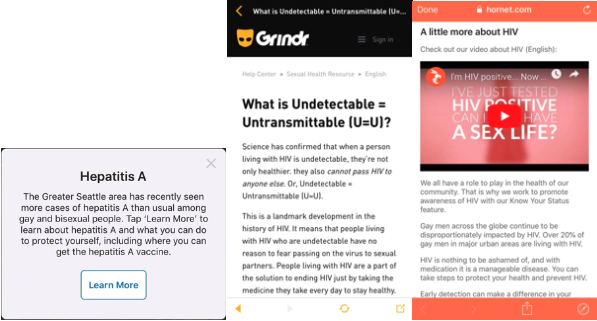}
\caption{Examples of HIV/STI Prevention Features on Grinder (left, middle) and Hornet (right)\label{F2}}
\end{figure}

Out of the 49 intimate platforms we examined, 13 were meant specifically for queer communities (11 specifically targeted at gay and bisexual men and 2 at lesbian and bisexual women). None of the platforms we reviewed were distinctly designed for trans people. The remaining 34 platforms were for general audiences, catering to heterosexual and homosexual connections, and 3 platforms were exclusively for heterosexual connections (eHarmony, Uniform Dating, and Waplog). Only queer-specific platforms (6) had explicit HIV disclosure options and allowed for filtering or searching based on HIV status. \hyperref[T2]{Table 2} shows the disclosure options for each platform. Growlr, Taimi, and Scruff allowed users to indicate that they were accepting of people living with HIV. Grindr, Hornet, Mr. X, Xtremboy, and Scruff, five platforms all of which are queer-specific, provide informational interventions with respect to HIV/STI prevention (See \hyperref[F2]{Figure 2} for examples). 8 dating apps mentioned HIV in their policies (5 queer-specific, 3 general). 4 dating apps allowed users to identify with an HIV/STI-relevant identity category, often labeled ``poz''.

\subsection{Policies}

None of the 49 intimate platforms we reviewed explicitly mention HIV in their terms of service. Four platforms expressly discuss HIV in their privacy policies (Grindr, Hornet, Scruff, and Mr. X), and four platforms mention HIV in platform safety policies (Planet Romeo, Tinder, BlackPeopleMeet, and Our Time). No platform engaged any of the legal implications of HIV. No platform engaged the public health surveillance of HIV.

Of the four platforms that expressly engage HIV in their privacy policies (Grindr, Hornet, Mr. X, Scruff), only two (Grindr and Hornet) explicitly prohibit sharing HIV information with third parties. By disclosing one’s HIV status on Mr. X and Scruff, users consent to the platform’s processing of that information. Grindr warns that HIV status disclosure on a user profile is effectively public information, however the platform does not share HIV status information with third party tracking, analytics, and advertising companies or service providers. Of all the platforms reviewed, Grindr’s privacy policy is the only one that devotes an entire section to HIV status, which is not particularly surprising given Grindr’s involvement in multiple controversies around sharing HIV information with third-parties~\citep{Fitzsimons2019,Singer2018}:

\begin{quotation}
\textit{``HIV Status. At the recommendation of HIV prevention experts and the community of Grindr users, we give you the option of publishing your health characteristics, such as your HIV status, in your Grindr community profile. Remember that if you choose to include information in your profile, that information will become public to other users of the Grindr App. As a result, you should carefully consider whether you want to disclose your HIV status. We do not share HIV status with any third-party service advertisers or third-party service providers other than companies that host data on our behalf (e.g., Amazon Cloud). In addition, we do not use HIV status for advertising purposes and do not share this information with advertisers.''}
\end{quotation}

According to Hornet’s privacy policies, they ``[do] not share any HIV status information with third parties unless required to do so by law''. Of the 49 platforms reviewed, Hornet was the only one to enable users to opt into receiving ``in-app reminders to undergo HIV tests and receive information on the location of nearby testing centers.'' On Hornet, a user’s HIV status ``is only searchable by users who have defined themselves as HIV positive.'' Scruff’s privacy policy highlights that ``there is no requirement to'' provide them with ``health details and whether part of the POZ (HIV positive) community (for example, in creating or updating your profile),'' and that by doing so, users ``are explicitly consenting to [Scruff’s] processing of [their] information.'' Mr. X’s privacy policy notes that HIV status information ``may be considered ‘special’ or ‘sensitive’ in certain jurisdictions,'' and that by providing this information, users ``consent to [Mr. X’s] processing that information''.

\section{Discussion}

\subsection{Prevention}

Platforms can act as an interventional tool to improve access to and perceptions of care for people living with HIV. Examples of HIV/STI prevention include a ``Last Tested Date'' section on a user’s profile and reminders to get tested for HIV/STIs. Some current platforms engage with HIV more critically by acknowledging that HIV is an issue its users should be aware through specific features. Hornet, for instance, provides its users with HIV-relevant educational material and resources for getting tested. Hornet also limits searching based on HIV status to people who themselves have chosen the HIV positive option, thereby limiting the possibility of HIV status-based discrimination. Hornet and Grindr can also provide reminders for users to get tested. Scruff allows users to choose from sex safety practices that include using condoms, using pre-exposure prophylaxis (PrEP), and/or treatment as prevention~\citep{warner2019signal}.

Due in large part to the history of HIV’s recognition as a medical condition, HIV has been generally classified as a ``gay man’s problem'' in North America---frequently (albeit almost as frequently unmarked) a white, cisgender gay man’s problem. This classification and framing acted to both separate normative society from the stigma associated with the condition and provide an avenue for activism by associating it with the most ``acceptable'' queer bodies: masculine, middle-class, cisgender and white~\citep{epstein1996impure}.

HIV has disproportionately impacted gay communities specifically, but transmission does not fit a neat pattern of being binarized tidily along sexuality. It is disproportionately prevalent in communities of color, appears in heterosexual relationships and lives, and risk of transmission follows societal vulnerability and marginalization — transgender women, particularly transgender women of color, are particularly overrepresented in diagnosis rates~\citep{clark2017diagnosed}. While the partial normalization of HIV---holding it outside the direct concerns of white, cisgender, heterosexual people, but embodying it in people who look ``just like them''---may have aided in assembling efforts to address the condition, the assumptions that it has created in who is at risk and who ``counts'' have been tremendous. One only has to look at the ethnographic work of Vivianne Namaste, who highlights how Montreal’s history of HIV, its recognition, and efforts at its prevention simultaneously elided the incidence rate amongst the Haitian community (which at one point had 65\% of reported AIDS cases) and lacked any advice or conception of susceptibility for women, particularly heterosexual or bisexual women~\citep{namaste2015oversight}.

Our platform analysis demonstrates that these same assumptions about vulnerability and risk are present in the design of intimate platforms. Generic platforms (i.e. those that cater to non-queer or broader, more heteronormative audiences) entirely do not consider, engage, or design for HIV while the platforms for queer---and more specifically gay men---do. Even within the group of thirteen queer-specific applications, neither of the two queer women-specific apps allowed for HIV disclosure, even though 23 percent of people with HIV in the US are women~\citep{cdc2019c}. Most, if not all, platforms dedicated to general audiences do nothing when it comes to HIV prevention, contributing to the knowledge gap for general audiences on sexual health, HIV-specific, and more. Because general audiences can go through online dating experiences without encountering HIV materials, platform designers allow these users to falsely believe that their sexual lives are excluded from important matters of sexual health. 

Our intent is not to suggest that HIV should be narrated as a problem for everyone; to ignore sexuality in the impact and risk of HIV transmission is an ahistorical mistake. But treating it solely as a ``gay man’s problem'' simultaneously elides differences in vulnerability and risk within gay communities and perpetuates the silence around transmission for other populations, particularly trans women of color and/or heterosexual people. In other words, it is not that HIV is not frequently a risk for gay communities, but that drawing a line between sexuality and risk perpetuates the more nuanced disparities in risk and the discourse that HIV transmission is not something anyone else has to think about.
As shown above, platforms can and have implemented prevention efforts through Last Tested Date and Testing Reminders features. Doing so more ubiquitously, rather than solely on gay male-specific platforms, may be helpful in normalizing prevention efforts like getting tested regularly and knowing one’s status.  Through opportunities like this, platform designers have the opportunity to promote HIV/STI prevention and care---an opportunity that is valuable precisely for its ability to normalize prevention efforts. This is not to say that such features are not without risks, particularly with regards to state surveillance, intervention and structural forces, which is our next topic of concern and discussion.

\subsection{Stigma \& Disclosure}

Designing for HIV is not as simple as including disclosure fields and status-based filtering or not. Allowing disclosure and filtering can protect people living with HIV from negative and sometimes harmful interactions, help filter out people who might discriminate against them, fight HIV stigma, and promote much-needed awareness. However, disclosure and filtering can also lead to discriminatory practices~\citep{hutson2018debiasing}, have potential for privacy unraveling~\citep{warner2018privacy}, and contribute to surveillance~\citep{fan2011sex,fan2012decentralizing}.

Destigmatizing HIV offers designers an opportunity to engage in the structural dimensions of how HIV operates in social life and can possibly allow us to better tap into social norms around the condition that ultimately improve other outcomes. For instance, humanizing people living with HIV could lead to more people getting tested, being open about their status, and being communicative with their sexual partners. Platforms have the power to shift social norms and destigmatize HIV at scale due to their pervasiveness throughout modern connections, but designers need to contest the ethical implications of destigmatizing HIV on these platforms, especially through current features such as HIV-status-based filtering and disclosure options.
	
Filtering and searching tools based on HIV status can be instrumental for people living with HIV to find others who are either seropositive or otherwise accepting of seropositive people. Additionally, filtering out those who might discriminate against them for their HIV status anyways allows people living with HIV to avoid awkward or even violent interactions with users who harbor problematic beliefs about people living with HIV. Conversely, HIV status-based filtering and searching tools have representational and allocational harms. First, it represents that there are particularly psycho-social characteristics incumbent with HIV status. These stereotypes play out in a variety of different ways such as the framing that people living with HIV engage in ``risky'' sexual behavior. Second, HIV status-based filtering can be used to structurally exclude HIV positive users from the opportunity to engage in intimate affiliation~\citep{hutson2018debiasing}. Platforms can and do provide users the ability to screen out other users who identify as ``Poz'' or disclose their HIV status. Not only do these design features facilitate exclusion, they may disincentivize HIV related disclosures to the extent that such disclosures can be weaponized by other users to exclude them as potential intimate affiliates.

Disclosure fields as a way to destigmatize HIV are similarly complicated in that they can inhibit and benefit people living with HIV. For one, encouraging users to disclose, regardless of their status, can create a healthier culture and discussion around HIV, possibly making talking about one’s status an acceptable and common practice of intimate engagement. On the other hand, disclosure can be used for a variety of problematic ends that harm seropositive users. Other users may discriminate against users who have disclosed their HIV status, choosing to ignore or disengage with them entirely. Disclosure may have unintended consequences and lead to more personal and violent outcomes. Due to laws in particular jurisdictions, failure to disclose one’s status to a partner can lead to prosecution and potentially incarceration. People living with HIV might also face physical and emotional threats for disclosing their status either publicly or privately.

Due to these complexities, designers of dating platforms must face the question of how can we destigmatize HIV without creating additional obstacles for people living with HIV? Platforms need to critically unpack the possible consequences of well-intentioned design choices, including HIV status-based filtering and HIV status disclosure fields. Of the platforms we reviewed, Scruff is the only one to provide for HIV disclosure without using an express ``HIV status'' field, allowing instead two disclosure options, Poz and Treatment as Prevention. ``Poz'' constitutes an association and identification with a community (e.g. ``I am a bear, daddy, poz''), while ``Treatment as Prevention,'' signals antiretroviral therapy (i.e. use of HIV medicines to treat HIV infection) and constitutes a link to sex safety practices.

\subsection{Surveillance \& Criminalization}

At the same time, given the questions of structural power and surveillance built into these platforms, we are leery of treating disclosure option design as the site of destigmatization and justice. Questions of privacy and stigma go wider than micro-interactions and touch on how HIV is seen and responded to societally and administratively. The dominant responses to HIV/AIDS ``center on adjusting the traditional levers of criminal and tort law, and of public health law, with its surveillance and disciplinary regimes that concentrate information and decision-making in the state''\citep[p.36]{fan2011sex}. Indeed, HIV continues to function as a ``vector for the exercise of state power and the invention of novel logics and techniques of government'', whereby ``[i]nfection with HIV virtually guarantees that a citizen will need to interact, either beneficently or coercively, with one or more state bureaucracies''\citep[p.255]{halperin2017war}.

The broader ecosystem of intimate platforms that we observed provided virtually no HIV-specific privacy information or protections for users living with HIV. Overall, both the platforms that account for HIV in their privacy policies and the platforms that enable disclosure but do not account for HIV in their privacy policies continue to place the risks and burden of surveillance, privacy, and disclosure on users with HIV. Grindr's ``HIV Status'' policy puts it clearly: ``Remember that if you choose to include information in your profile, that information will become public to other users of the Grindr App.'' By surfacing this as a risk we do not mean to suggest that users lack agency---merely that the agency to choose between a range of options can be influenced by how those options are bounded and made available in addition to the affordances and norms that platform design provides. That a user makes information public does not mean that ``consumable by all'' is the framework of disclosure that they have in mind ~\citep{wittkower2016lurkers}.

While some intimate platforms are working towards promoting HIV disclosure, prevention, and destigmatization, they are also failing to grapple with privacy implications of HIV and their responsibility in ensuring it. People living with HIV are already vulnerable and bear the weight of HIV disclosure’s downstream consequences. By continuing to offload the burdens and risk on those with HIV, platforms are likely contributing to issues of nondisclosure as well as HIV testing. Research shows that privacy fears can result in the non-disclosure of HIV status information within close personal relationships\citep{derlega2002perceived,zea2003asking,derlega2004reasons}.

In this context, proposals to design for HIV disclosure that do not consider the wider structural implications of surveillance are concerning. The focus of most research into HIV and online dating in HCI on micro-interactions and enabling trust and certainty between users elides the implications that providing this data to a platform outside user control has and the way that this data can be used to control. This is not an abstract risk; just this year, Grindr (the platform under study) has been the subject of scrutiny by the U.S. government over its Chinese ownership, due to fears that the Chinese government might access and copy Grindr’s data around HIV disclosure for the purpose of domestic policing and control~\citep{Fitzsimons2019}. If we are designing to enable HIV disclosure, are we working to improve stigma associated with disclosure - or are we enabling new forms of control and surveillance?

In the United States today, intimate platforms operate within 29 states that have HIV-criminal laws, which include laws that target sex/nondisclosure of HIV-positive status, sex work, exposure to bodily fluids, needle-sharing, sex work, and blood/organ/semen donation,  9 states that have sentencing enhancements applicable to people living with HIV who commit an underlying assault crime, and 24 states that have prosecuted people living with HIV under non-HIV-specific general criminal laws\citep{chiv2019}. Here, the design of intimate platforms cannot removed from the reality of laws that criminalize HIV, particularly HIV non-disclosure. 

People living with HIV in U.S. with HIV-specific criminal laws must disclose their HIV status to sexual partners. Generally, ``disclosure and consent'' is an affirmative defense, whereby a person can avoid criminal and civil liability if they disclose their serostatus and their sexual partner voluntarily consents to sexual activity with knowledge of that serostatus~\citep[p.1101]{lehman2014prevalence}. Many of the laws that criminalize HIV non-disclosure do not provide guidance as to what methods of disclosure and consent are enough to avoid prosecution and conviction~\citep{mccallum2014criminalizing}. No court or legislature has affirmatively stated whether verbal disclosure and consent are necessary under criminal HIV transmission statutes. Furthermore, non-verbal communication online create uncertainty as to whether there is sufficient disclosure and consent to remove criminal liability for HIV-positive individuals. Both disclosure and consent can be ambiguous or misunderstood, a problem that is complicated by the design and widespread use of mobile dating and hookup platforms.

It remains unclear what constitutes appropriate disclosure and informed consent in the context of intimate platforms, such as HIV disclosure fields on user profiles or other communication in a profile's free form text sections (e.g. ``+'', ``Poz'', ``undetectable''). Although some intimate platforms afford HIV-positive users the ability to disclose their serostatus in new ways, no court or legislature in the U.S. has answered whether disclosing HIV status on an intimate platform is enough to achieve informed consent and avoid criminal and civil liability. Yet many people living with HIV also use records of conversations on intimate platforms as a means of protection. For example, people disclose their status and use that record as a way to protect themselves from future allegations of non-disclosure. This ambiguity and incumbent legal risk places significant responsibility and pressure on HIV users. Research shows that fears around rejection, self-blame, criminalization, and privacy can result in the non-disclosure of HIV status information within close personal relationships~\citep{derlega2004reasons,zea2003asking,derlega2002perceived}. Privacy concerns around HIV disclosure are often associated with the need to protect one’s self from HIV related stigma~\citep{adam2011hivstigma,serovich2003reasons,greene2003privacy}. As more and more people use platforms to meet intimate partners, the historical failure of HIV criminalization law to understand how disclosure and consent are negotiated in practice becomes all the more apparent. 

It might seem from this that designers and developers are trapped in an impossible situation---disclosure to protect users simultaneously produces the possibility of structural harms for those disclosing. While we urge designers to take both needs seriously, we do not consider it impossible; in fact, there is a range of work within queer theory and technology that not only articulates this tension of privacy, disclosure and the reuse of data but suggests queer forms of resistance to it. Writing more broadly, Brian Schram highlights the way that the increasing possibilities of ``big data'' and its attendant surveillance structures ``constitute an undoing of Queerness as a radical political injection''~\citep[p.611]{schram2019accidental}, advocating a politics of melancholia that features ``a haunting of archives: an insertion of the dead weight of our collective memory as Queer persons into the growing catalog of our digital information''. In other words, Schram suggests the deliberate incorporation of masses of false data, profile and traces into data stores in order to render ambiguous the truth of any presence and provide cover for those queer persons existing within the platform(s) data highlights. What would this look like in the case of dating platforms? What are the possibilities raised by incorporating a deluge of false accounts, doppelgangers and doubles, not as a deception of the platform or its users, but against state forces examining the database?

More broadly, we might see possibilities for the future through practices in the past. In how queer communities responded to HIV disclosure and protection protocols during the 1980s and 1990s, David Halperin has articulated the way that gay communities worked to articulate norms that balanced risks, trust, and vulnerability in the absence of structural norms, that ``it is gay men themselves who have continued to define, and to redefine, the limits of safety through an ongoing history of sexual experimentation and mutual consultation, and who have thereby produced, over time, workable compromises and pragmatic solutions that balance safety and risk'' \citep[p.207]{halperin2016biopolitics}. Rather than taking universalized, top-down approaches to platform design for all, we might instead seek to work up and to create a diverse range of spaces that challenge the ease of surveillance built into large-scale platforms and afford individual users more agency in establishing those compromises and solutions and engaging in that consultation. 

\section{Conclusion}
As HCI researchers and designers, we continue to push the boundaries of what is technologically possible but doing so requires us to first ask whether platform design is even an appropriate intervention in a given situation~\citep{keyes2019mulching,baumer2011implication,suchman2011anthropological}. The current model of platform design for HIV cannot continue, as it is too closely tied to the collection and commodification of highly sensitive personal data. However, reimagining intimate platform design provides the social computing community an opportunity to intervene in the social norms around HIV and HIV disclosure in a manner that could unburden the weight of criminalization without centralizing the surveillant arms of the state.

We envision a future of dating platforms that does not force people living with HIV to sacrifice surveillance for intimate experiences. Because of their entanglements with sex and romance, intimate platforms need to take on more responsibility in the sexual health and data privacy of their users. Drawing from our analysis and our own lived experiences, we recommend platform-level changes, changes in platform, and mechanisms to prevent platforms from knowing their users’ statuses. First, platforms should make explicit to their users the consequences of storing sensitive, personal information like HIV status and their documentation processes. Next, they should also implement policies that manage how data are stored when users delete their accounts and protect these data from third-party consumers. Finally, ownership of user’s data should belong to the users themselves, rather than the platforms. Users should be able to pass along their information to other users without the platforms tracking it.

HIV is a medical condition, but its eradication requires not just technical, or even sociotechnical, but sociopolitical solutions. Indeed, the ways in which designers and policymakers frame HIV is an inherently political decision, one that will impose the contours and boundaries of our response. The social computing community cannot do nothing, but it also must resist the desire to do everything. Designing user interfaces and platform policies to account for HIV will require a rigorous analysis of possible outcomes and consequences as well as a bedrock commitment to centering the voices and experiences of those impacted by HIV and the state’s responses to it. Our commitments must account for the ways pathology and power intertwine to subjugate and otherize impacted communities at home and abroad.

Designing intimate platforms to unburden the risks of extant criminal and civil sexual regulations runs the risk of re-entrenching the status quo and its incumbent inequalities and power relations~\citep{dombrowski2016social,irani2010postcolonial,light2011hci,bardzell2010feminist}. The social computing community must ground its efforts to design for HIV in clear political commitments to decriminalizing HIV and decentralizing power and information from the state. We must strive to unburden the weight of surveillance and incarceration on vulnerable and marginalized communities and work towards offloading the significant social and legal risks and pressures for people living with HIV. Moreover, our commitment to designing for HIV must not exclude nor obfuscate our capacity for direct action within and outside of the realms of design and research. This means fighting for the rights, dignity, and safety of people living with HIV in the streets and in the halls of local, national, and international political, legislative, and executive bodies. 

Our instinctual response to the failed and violent efforts of HIV criminalization and surveillance should not be ``there’s an app for that'', but rather ``there’s a zap for that!''. That is, the practice of designing for people with HIV should be a ``critical technical practice'' \citep{agre1997lessons}, undertaken with a mindset that sits uneasily between and is cognizant of both individual and structural power and consequence. Pioneered by the American gay liberation movement, a zap or ``zap action'' is a political action of direct and persistent public confrontation. Whether shouting down public figures or smashing pies into the faces of evangelicals, zaps aim to disrupt and disturb persons and institutions of authority to effect change ~\citep{Cohen2018}. In the words of AIDS Coalition to Unleash Power’s (ACT UP) ``New Member Packet'':

\begin{quotation}
\textit{``Zaps are a method for ACT UP members to register their disapproval of and anger toward the zap target. Zaps usually have more specific targets than actions. Because of this focus, numerous zapping techniques have been developed. ACT UP zaps individuals or organizations by: sending postcards or letters; invading offices and distributing fact sheets; sending (lots and lots of) faxes; picketing; outraged (and sometimes outrageous) phone calls. The more zappers who zap the zappee the better the zap.''}
\end{quotation}

A critical approach to designing for HIV requires the contesting of histories of incarceration, stigmatization, and surveillance and the ways in which the state exerts power and domination through its medicolegal levers of criminal law and public health surveillance. Intimate platform design should not only work to reduce the prevalence and stigma of HIV, but also to contest historic and present power imbalances and injustices between users, platforms, and the state.

\begin{acks}
We dedicate this paper to the radical history of the AIDS Coalition to Unleash Power (ACT UP) and to all of the souls we've lost and continue to lose to HIV/AIDS. We would like to thank Mary Fan, Sean Munson, and Julie Kientz for valuable conversations and feedback, and Margret Wander and Margaret Hopkins for their ongoing care and support.
\end{acks}

\begin{funding}
This research was partially funded by a Microsoft Ada Lovelace Fellowship.
\end{funding}

\bibliographystyle{SageH}
\bibliography{main}

\end{document}